\documentclass{INTERSPEECH2023}
\usepackage[dvipsnames]{xcolor}
\usepackage[T1]{fontenc}
\usepackage{multirow}

\interspeechcameraready

\title{Learning Emotional Representations from Imbalanced Speech Data for Speech Emotion Recognition and Emotional Text-to-Speech}
\name{Shijun Wang$^1$, Jón Guðnason$^2$, Damian Borth$^1$}
\address{
  $^1$University of St.Gallen, Switzerland\\
  $^2$Reykjavik University, Iceland}
\email{shijun.wang@unisg.ch, jg@ru.is, damian.borth@unisg.ch}

\begin{document}

\maketitle
 
\begin{abstract}
Effective speech emotional representations play a key role in Speech Emotion Recognition (SER) and Emotional Text-To-Speech (TTS) tasks.
However, emotional speech samples are more difficult and expensive to acquire compared with Neutral style speech, which causes one issue that most related works unfortunately neglect: imbalanced datasets.
Models might overfit to the majority Neutral class and fail to produce robust and effective emotional representations.
In this paper, we propose an Emotion Extractor to address this issue.
We use augmentation approaches to train the model and enable it to extract effective and generalizable emotional representations from imbalanced datasets.
Our empirical results show that 
(1) for the SER task, the proposed Emotion Extractor surpasses the state-of-the-art baseline on three imbalanced datasets;
(2) the produced representations from our Emotion Extractor benefit the TTS model, and enable it to synthesize more expressive speech.
\end{abstract}


\section{Introduction}

Speech Emotion Recognition (SER) and Emotional Text-To-Speech (TTS) are two important speech-processing tasks.
SER studies how to recognize emotional information from speech, while Emotional TTS studies how to generate speech with rich emotional styles given texts.
Interest in both tasks is on the rise because they enable the recent machine learning models to better understand and mimic human emotions. 
The success of these two tasks shares one key role: the effectiveness of speech emotion representations \cite{Latif2021SurveyOD, Triantafyllopoulos2022AnOO}.
Specifically, emotional information carried by meaningful representations benefits SER, meanwhile, it renders expressive emotional speech production. 

However, one challenge for SER and Emotional TTS is the lack of data, which has not been taken into account in most related approaches \cite{Kim2022ImprovingSE, Liu2022DiscriminativeFR, Lei2022MsEmoTTSME, Im2022EMOQTTSEI, Wang2023FinegrainedEC}.
In reality, emotional speech data is more difficult and expensive to acquire compared to Neutral style speech,
which results in a common issue that the speech data distributions are highly skewed toward the Neutral emotion class.
Therefore, when trying to extract emotional representations, we need to account for imbalanced datasets.

Data augmentation is a common strategy to alleviate the imbalanced data issue,
by reducing the bias towards the majority class and increasing the robustness of the model. The authors from \cite{Chatziagapi2019DataAU, Latif2020AugmentingGA, Wang2022GenerativeDA} apply GANs to augment speech and improve the SER performance on imbalanced datasets. 
In order to increase the performance of TTS models on small datasets, recent studies propose to augment small corpora with synthesized data, either via voice conversion \cite{Huybrechts2020LowResourceET, Ribeiro2022CrossSpeakerST} or a large text-to-speech system \cite{Hwang2020TTSbyTTSTD}. 
However, most of these TTS works focus on synthesizing augmented speech, rather than using augmentation to achieve more effective speech emotional representations.
Meaningful emotional representation is crucial for expressive emotional TTS and can be utilized to affect speech synthesis effectively.
The authors in \cite{Wang2018StyleTU} learn style tokens to represent emotional information from reference speech, then these tokens are used to transfer the emotion to the synthesized speech.
RFTacotron \cite{Lee2018RobustAF} is an extended work of \cite{Wang2018StyleTU}. 
Instead of using one single token, RFTacotron applies a sequence of vectors to reflect the emotional style of the reference. 
Such representations provide fine-grained and robust emotional-style control on the synthesized speech.
However, it might be difficult for these models to learn effective emotional representations from imbalanced data, 
because the models might overfit to Neutral emotion, which makes the generated speech less expressive.

Mixup \cite{Zhang2017mixupBE} is a popular way to address imbalanced data issues.
It augments data by linearly interpolating input samples and the corresponding labels.
The effectiveness of Mixup has been widely demonstrated.
In \cite{Meng2021MixSpeechDA}, the authors use Mixup to improve speech recognition performance on small datasets. 
Mixup is also applied in \cite{Zhang2022ContrastivemixupLF} to enhance model generalization for speaker verification tasks.
Despite the success of Mixup augmentation directly on speech (linear interpolation of two raw speech samples, i.e. raw-level Mixup), the authors in \cite{Kang2022LMixAL} argue that samples created by Mixup are likely to be very unrealistic in terms of speech production, 
and the effect could be limited due to its linear nature. 
As a result, \cite{Kang2022LMixAL} applies Mixup to latent space rather than raw inputs, i.e. Mixup on latent representations.
Such latent-level Mixup has been explored in other domains as well \cite{Beckham2019OnAM, Sun2020MixupTransformerDD}. 

In this work, we propose an augmentation approach that can effectively learn speech emotion representations from imbalanced data. 
Such representations can be used for both SER and Emotional TTS tasks.
Specifically, we combine both raw- and latent-level Mixup to gain benefits from both sides.
In raw-level Mixup, the model is exposed to a wider variety of input data, and learns to identify the underlying structure in the data, which will likely lead to a more robust and generalizable feature space.
On the other hand, Mixup on latent space is exposed to wider information at the level of intermediate activations, which results in a more expressive feature space. 
Consequently, such manipulation allows us to extract more complex and abstract features of the data, which can be especially beneficial for tasks requiring high-level understanding, 
To sum up, the combination of these two types of Mixup augmentations has the potential to lead to robust and effective emotional representations.

Additionally, since we can consider raw- and latent-level Mixup two types of augmentation, we force the emotional representations from these two augmentations to attract each other. 
This is motivated by recent successful Contrastive Learning methods \cite{Oord2018RepresentationLW, Chen2020ASF, Caron2020UnsupervisedLO, Scheibenreif2022SelfsupervisedVT}. 
 By forcing the two representations to be similar, the model is encouraged to produce consistent representations for the mixed instances, regardless of whether the mixing is performed at the raw level or at the latent level. 
 Moreover, this consistency might prevent the model from being overly dependent on specific values or features that are likely to be different between the training and test sets in imbalanced datasets.
One thing that is worth mentioning is that although we do not feed negative samples as most of the Contrastive Learning methods, we can still learn discriminative representations, since the standard classification loss (cross-entropy) is still applied with the help of emotion labels. 

We summarize our contributions as: 
(1) we propose an augmentation approach that can learn effective emotion representations from imbalanced datasets.
(2) The learned representations can be applied to both SER and Emotional TTS tasks.
(3) Our experimental results demonstrate that our SER and Emotional TTS models benefit from the learned representations and outperform state-of-the-art models.




\section{Methodology}

This section describes how we learn emotional representation  from speech data and how they are used for SER and Emotional TTS.


\subsection{Emotional Representation Learning}

\begin{figure}[t]
  \centering
  \includegraphics[width=0.85\linewidth]{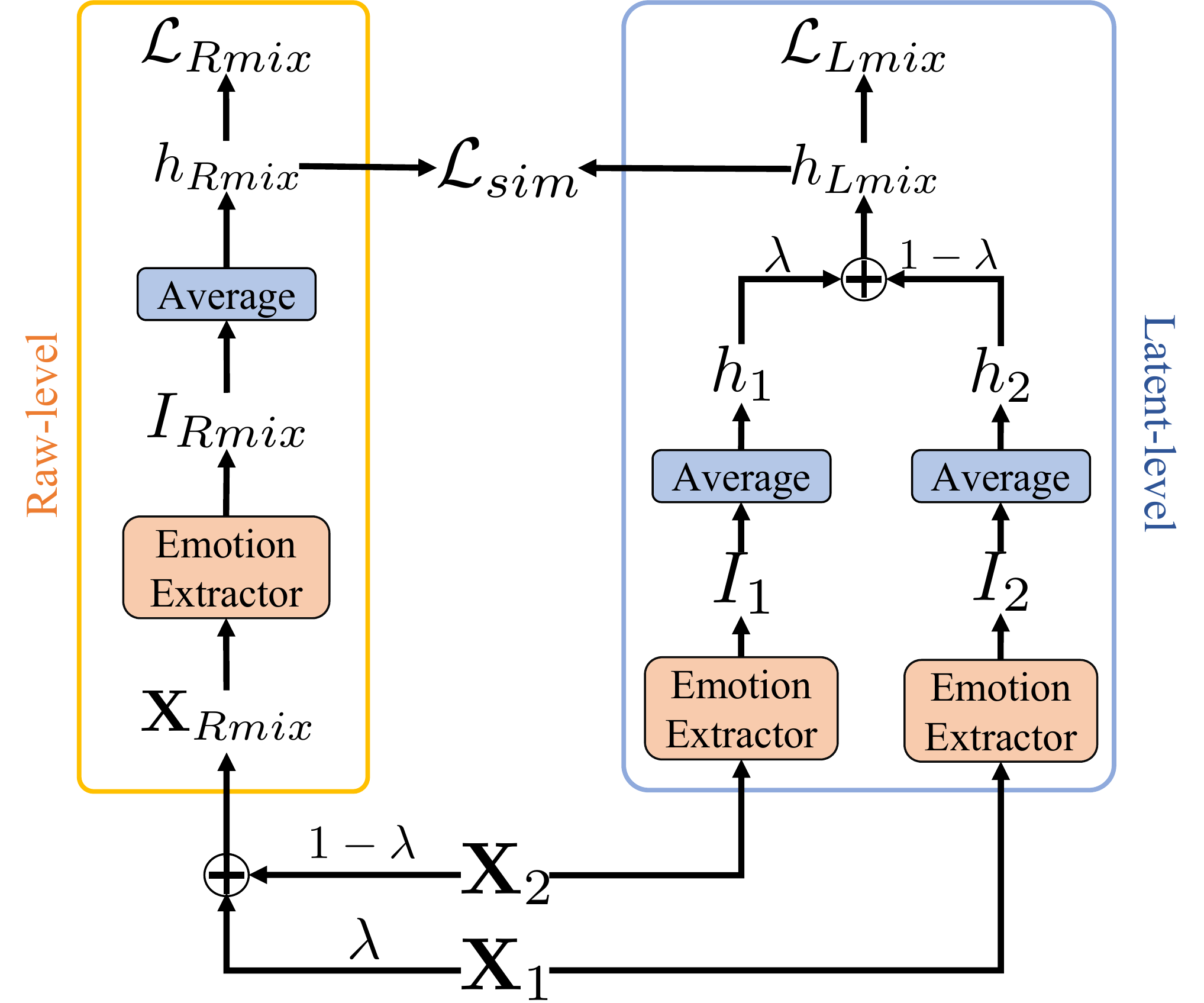}
  \caption{The training of the Emotion Extractor. $\mathbf{X}_1$ and $\mathbf{X}_2$ are two speech samples. 
  The left path is raw-level Mixup. 
  The mixture $\mathbf{X}_{Rmix}$ is sent to the Emotion Extractor to produce the emotion representation $\boldsymbol{I}_{Rmix}$. $h_{Rmix}$ is an averaged vector. 
  The right path is latent-level Mixup, we first get individual representations $\boldsymbol{I}_{1}$ and $\boldsymbol{I}_2$ from $\mathbf{X}_1$ and $\mathbf{X}_2$. 
  After average, we perform Mixup on vectors $h_{1}$ and $h_{2}$ to acquire the mixture $h_{Lmix}$. 
  Finally, raw- and latent-level losses $\mathcal{L}_{Rmix}$ and $\mathcal{L}_{Lmix}$ are two weighted cross entropy losses, while $\mathcal{L}_{sim}$ is to encourage $h_{Rmix}$ and $h_{Lmix}$ to be similar.}
  \label{fig:EE}

\end{figure}

We train an Emotion Extractor to extract emotional representations. The pipeline for the training is shown in Fig.~\ref{fig:EE}. 
We have two speech samples $\mathbf{X}_1$ and $\mathbf{X}_2$ as inputs. 
The left is raw-level Mixup (standard Mixup). We perform this augmentation on the pair ($\mathbf{X}_1$, $\mathbf{X}_2$) to get the mixture $\mathbf{X}_{Rmix}$:
\begin{equation}
\mathbf{X}_{Rmix} = \lambda \mathbf{X}_{1} + (1-\lambda) \mathbf{X}_{2}, 
\end{equation}
where $\lambda$ is sampled from Beta distribution $Beta(1,1)$.

Raw-level mixed output $\mathbf{X}_{Rmix}$ is then sent to our Emotion Extractor to produce emotion representation $\boldsymbol{I}_{Rmix}$. Afterward, we average the sequence $\boldsymbol{I}_{Rmix}$ into a vector $h_{Rmix}$. The standard Mixup loss is further applied to it:
\begin{equation}
\mathcal{L}_{Rmix} = \lambda \text{CE}(h_{Rmix}, y_{1}) + (1-\lambda) \text{CE}(h_{Rmix}, y_{2}), 
\end{equation}
and CE($\cdot, \cdot$) represents Cross Entropy loss, $y_{1}$ indicates the emotion category label of $\mathbf{X}_{1}$, while $y_{2}$ indicates the label of $\mathbf{X}_{2}$.

On the right side, we perform latent-level Mixup augmentation. 
The same speech inputs $\mathbf{X}_1$ and $\mathbf{X}_2$ are first independently fed to our Emotion Extractor to produce individual emotion representations $\boldsymbol{I}_{1}$ and $\boldsymbol{I}_{2}$. 
Both representations are averaged into $h_{1}$ and $h_{2}$. 
Latent-level Mixup is then applied to them:
\begin{equation}
{h}_{Lmix} = \lambda h_{1} + (1-\lambda) h_{2}.
\end{equation}
Similar to $\mathcal{L}_{Rmix}$, we apply latent-level Mixup loss on $h_{Lmix}$:
\begin{equation}
\mathcal{L}_{Lmix} = \lambda \text{CE}(h_{Lmix}, y_{1}) + (1-\lambda) \text{CE}(h_{Lmix}, y_{2}).
\end{equation}

Furthermore, as we mentioned before, in order to encourage the model to produce generalizable emotional representations better, we force the outputs of the raw- and latent-level Mixup augmentations to be similar.
Therefore, we employ an extra loss $\mathcal{L}_{sim}$ on $h_{Rmix}$ and $h_{Lmix}$,
\begin{equation}
\label{L_sim}
\mathcal{L}_{sim} = - ~ \text{dot}(h_{Rmix}, h_{LMix}),
\end{equation}
where dot($\cdot, \cdot$) indicate the dot product between two vectors.

Finally, we train our Representation Extractor with the total loss:
\begin{equation}
\label{L_total}
\mathcal{L}_{total} = \mathcal{L}_{Rmix} + \mathcal{L}_{Lmix} + \mathcal{L}_{sim}.
\end{equation}

\subsection{SER and Emotional TTS}


We use two different architectures of the Emotional Encoder for the SER and Emotional Recognition tasks, since we want to make a fair comparison between our models and baselines from different tasks.

\subsubsection{SER Task}

Since our Emotion Extractor can be directly used for the SER task (Fig.~\ref{fig:EE}). We follow \cite{Wang2022GenerativeDA} and use VGG19 \cite{Simonyan2015VeryDC} as our Emotion Extractor for the SER task.
To adapt VGG19, we slightly modify our training pipeline in Fig.~\ref{fig:EE}. Specifically, we remove all Average operations since VGG19 gradually downsamples the input via Maxpooling.
VGG19 has multiple Convolutional layers and two fully-connected layers, thus, we refer to the last Convolutional layer's output as emotional representation $\boldsymbol{I}$, and the last fully-connected layer's output as vector $h$.

\subsubsection{Emotional TTS Task}

We use RFTacotron \cite{Lee2018RobustAF} as our backbone TTS to convert the phonemes to speech, given emotional representations. 
RFTacotron uses a Reference Encoder (multiple Convolutional layers and one GRU layer) to extract emotional representations from Mel-Spectrogram, and all modules are trained together.
In our work, the architecture of our Emotion Extractor is identical to that of the Reference Encoder.
Additionally, for the rest parts, we maintain the original model configuration.
Therefore, the only difference is that during the training of the TTS model, we use a pre-trained and frozen Emotion Extractor to provide emotional representations.

\begin{figure}[t]
  \centering
  \includegraphics[width=0.61\linewidth]{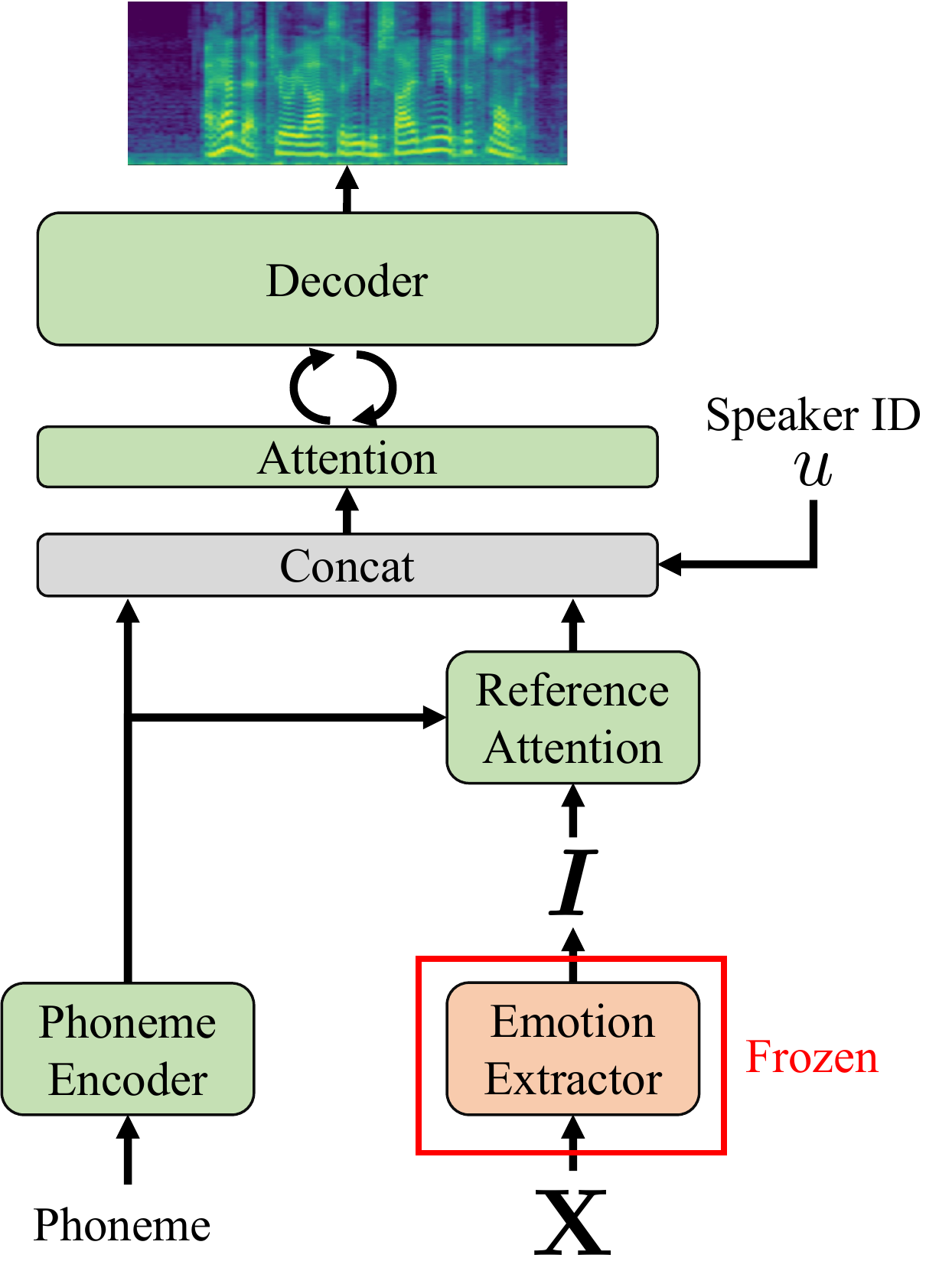}
  \caption{The training of our TTS model, a pre-trained Emotion Extractor is frozen and combined to provide emotional representations from the input speech $\mathbf{X}$.}
  \label{fig:TTS}

\end{figure}

The training pipeline is visualized in Fig.~\ref{fig:TTS}. We only give a short description of each module here but refer the readers to the original paper \cite{Lee2018RobustAF} for a more in-depth description.
The Phoneme Encoder is to process the input phoneme sequence. 
We use an attention module (Reference Attention) to align the phoneme sequence and the emotional representation $\boldsymbol{I}$, since their lengths are different.
Additionally, speaker ID $u$ is mapped into an embedding to represent different speaker characteristics. 
Afterward, the speaker embedding, phoneme, and emotional representation are concatenated together and sent to the decoder.
With another attention module, the decoder aggressively generates the final Mel-Spectrogram.

\subsection{Training Details}

\noindent 
\textbf{SER}: For the SER task, we train our Emotion Extractor with Adam optimizer for 30k iterations, and set the learning rate to 1e-6.

\noindent 
\textbf{Emotional TTS}: For this task, we first train our Emotion Extractor, then the pre-trained Emotion Extractor is frozen during the training of the TTS model. 
We train our Emotion Extractor for 30k iterations, and set the learning rate to 1e-6. The TTS model is trained using a 1e-4 learning rate over 600k iterations. Adam optimizer is used for both steps.

\section{Experiments}

\subsection{Experiment Setup}

\subsubsection{Dataset}

\noindent 
\textbf{SER}: EmoV-DB \cite{Adigwe2018TheEV}, IEMOCAP \cite{Busso2008IEMOCAPIE}, and CREMA-D \cite{Cao2014CREMADCE} are three datasets used for our SER task. 
EmoV-DB contains 4 speakers and five emotions (Angry, Sleepy, Neutral, Disgusted and Happy), the overall speech samples are around 7 hours.
IEMOCAP has 5 speakers, as many other SER works \cite{Latif2020AugmentingGA, Wang2022GenerativeDA}, we only focus on four classes (Angry, Sad, Neutral and Happy), resulting in 5531 speech utterances of about 7 hours total duration. 
CREMA-D contains 6 emotions (Angry, Sad, Neutral, Disgusted, Fear and Happy), amounting to 7742 speech utterances of about 4 hours. 
Furthermore, for all datasets, in order to simulate the data imbalance issue, we follow \cite{Wang2022GenerativeDA} and randomly remove 80\% of each class except Neutral.


\noindent 
\textbf{Emotional TTS}: For the Emotional TTS task, we employ ESD dataset \cite{Zhou2020SeenAU}. ESD has 5 emotions (Neutral, Angry, Sad, Happy and Surprised). In our experiments, we only use the data from 10 English speakers, with each speaker reading 350 utterances for each emotion. 
To simulate the data imbalance issue, during the training, for each speaker, we only keep 10 utterances (randomly selected) for each emotion class except Neutral.

\noindent 
\textbf{Data Preprocessing}: All speech samples from the aforementioned datasets are resampled to 16kHz. We use the same acoustic features as in \cite{Wang2022GenerativeDA}. Specifically, to extract Mel-Spectrograms from waveform files, we use a 50-millisecond window and a 50 percent overlap ratio, with 128 Mel Coefficients.

\subsubsection{Baselines}

\noindent 
\textbf{SER}: The baseline approach we use for the SER task is from \cite{Wang2022GenerativeDA}, where the authors use a GAN-based model to perform automatic augmentation to enhance the SER performance for imbalanced datasets. 
We follow the same experimental setup and model configurations to ensure a fair comparison.

\noindent 
\textbf{Emotional TTS}: We employ RFTacotron \cite{Lee2018RobustAF} as our TTS baseline model. RFTacotron uses a Reference Encoder to extract a sequence of vectors as the emotional representation from reference speech, this representation is further applied to influence speech generation. 
In our experiments, we replace the outputs of the Reference Encoder with the representations extracted from our Emotion Extractor. 
If our representations are more effective, then the synthesized speech should be more expressive. 

\subsubsection{Evaluation Setup}

\noindent 
\textbf{SER}: We conduct 5-fold cross-validation on all SER datasets to fully utilize the data and produce low-biased results. 
In this work, like many others \cite{Latif2020AugmentingGA, Wang2022GenerativeDA}, we report Unweighted Average Recall (UAR) for imbalanced datasets. 
Precision would be of less interest in our case, since we focus on False Negative (when minority samples are misclassified as a majority), rather than False Positive (when majority samples are misclassified as a minority).



\noindent 
\textbf{Emotional TTS}: We use HiFi-GAN \cite{Kong2020HiFiGANGA} to convert generated Mel-Spectrograms from the TTS models to waveforms. 
To perform subjective tests, for each emotion, we randomly select 15 unseen text utterances and use another 15 unseen speech samples as references. 
Thus, 75 speech samples can be synthesized from each model.
We conduct subjective evaluations on Amazon Mechanical Turk, and 20 subjects participate in the subjective evaluations. Demo page can be found at
\url{https://anonymous.4open.science/w/INTERSPEECH2023-F8C4/}

\subsection{SER Results}
\label{section:SER_Results}

\subsubsection{Ablation Study}

We first perform experiments to investigate the impact of each loss from Eq.~\ref{L_total}. The UAR results are listed in Tab.~\ref{tab:ablation_study}. 
As we can see, latent-level Mixup augmentation performs better than raw-level Mixup (the first two rows). 
The SER performance improves when we combine both Mixup augmentations, demonstrating that the model profits from both types of mixing and produces more effective and robust emotional representations.
After we add $\mathcal{L}_{sim}$ (Eq.~\ref{L_sim}), the model can exceed all previous models (the last row). Such results demonstrate that forcing the representations obtained from raw- and latent-level Mixup to be similar, improves generalization performance and mitigates the impact of train-test distribution shift, as we claimed before.

According to the result of the Ablation Study, in the rest experiments, we only use models trained with $\mathcal{L}_{total}$.

\begin{table}[t]
  \small
  \caption{Ablation Study (UAR Results) on an 80\%-reduced imbalanced IEMOCASP dataset.}
  \label{tab:ablation_study}
  \centering
    \begin{tabular}{ c | c | c | c || c }
    \toprule
     Training with               & Angry   & Sad    & Happy    & Average                            \\ 

    \midrule
    only $\mathcal{L}_{Rmix}$          & 54.86        & 55.93   & 50.06     & 53.62        \\ 

    only $\mathcal{L}_{Lmix}$          & 57.12        & 59.88   & 53.50     & 56.83        \\ 

    $\mathcal{L}_{Rmix} + \mathcal{L}_{Lmix}$          &59.74        & 60.09   & 54.56     & 58.13        \\ 

    $\mathcal{L}_{total}$          & \textbf{61.33}        & \textbf{62.73}   & \textbf{58.82}     & \textbf{60.29}        \\

    \bottomrule
    \end{tabular}
\end{table}

\begin{table}[t]
  \small
  \caption{UAR results on three imbalanced datasets. The emotion classes except Neutral of all datasets are reduced by 80\%.}
  \label{tab:SER}
  \centering
    \begin{tabular}{ c | c c | c c  | c  c  }
    \toprule
                            & \multicolumn{2}{c|}{EmoV-DB}                       & \multicolumn{2}{c|}{IEMOCAP}           & \multicolumn{2}{c}{CREAM-D}              \\ 
    \midrule
    Emotion                 & \cite{Wang2022GenerativeDA}     &  Ours            & \cite{Wang2022GenerativeDA}           & Ours     & \cite{Wang2022GenerativeDA}           & Ours  \\

    \midrule
    \midrule
    Angry                   &87.6 &\textbf{90.5}            & \textbf{62.8}                & 61.3            & 73.9           & \textbf{74.1}              \\ 

    \midrule

    Sad                     & - & - & 58.1                & \textbf{62.7}            & 44.2           &\textbf{46.9}              \\ 

    \midrule

    Happy     &84.6 &\textbf{86.9}  & 53.7                & \textbf{58.8}            & 39.2           & \textbf{42.6}              \\ 

    \midrule

    Sleepy     &79.8 &\textbf{83.6} & -                & -            & -           & -              \\ 

    \midrule

    Disgusted  &82.0 &\textbf{84.1} & -                & -            & 44.8           & \textbf{44.8}              \\ 
    \midrule
    Fear  & - & - & -                & -            & 34.3           & \textbf{41.0}              \\ 
    \midrule
    \midrule
    Average    & 83.5 & \textbf{86.3} &  58.2                & \textbf{60.9}            & 47.3           & \textbf{49.9}              \\

    \bottomrule
    \end{tabular}

\end{table}

\subsubsection{Comparison with the Baseline}

We compare our model with the baseline \cite{Wang2022GenerativeDA} and list the UAR results in Tab.~\ref{tab:SER}. 
The left is the results from the EmoV-DB dataset, our model outperforms the baseline for each emotion class, and is around 3\% better on average.
The results of the IEMOCAP dataset are in the middle.
Despite the baseline being better than our model for Angry emotion, our model performs better than the baseline for the other two classes by at least 2\%, 
which yields an average UAR performance that is about 2\% higher than the baseline.
The UAR results of CREAM-D are listed on the right side.
CREAM-D is more challenging than EmoV-DB and IEMOCAP, since it is smaller and contains more emotion classes and speakers. 
Nevertheless, similar to EmoV-DB and IEMOCAP, our model significantly exceeds the baseline and outputs a performance that is almost 3\% greater than the baseline.
 It is worth mentioning that the Angry class achieves much higher UAR than the other classes.  
 A likely reason for this is that speakers read this emotion louder than other emotions, which makes the Angry class easier to recognize.


From the results of three imbalanced datasets, we observe that the emotional representations extracted by our model are more generalizable and bring benefit for the SER performance on imbalanced datasets.

\subsection{Emotional TTS Results}

\subsubsection{Preference Test}

We carry out subjective A/B preference tests to evaluate the models' abilities of emotional expressiveness.
Given a pair of synthesized speech samples and one reference, subjects are asked to select the one that conveys emotion more clearly and has a speaking style close to the reference. 
If there is no detectable difference, they should choose "Same".

\begin{figure}[t]
  \centering
  \includegraphics[width=0.9\linewidth]{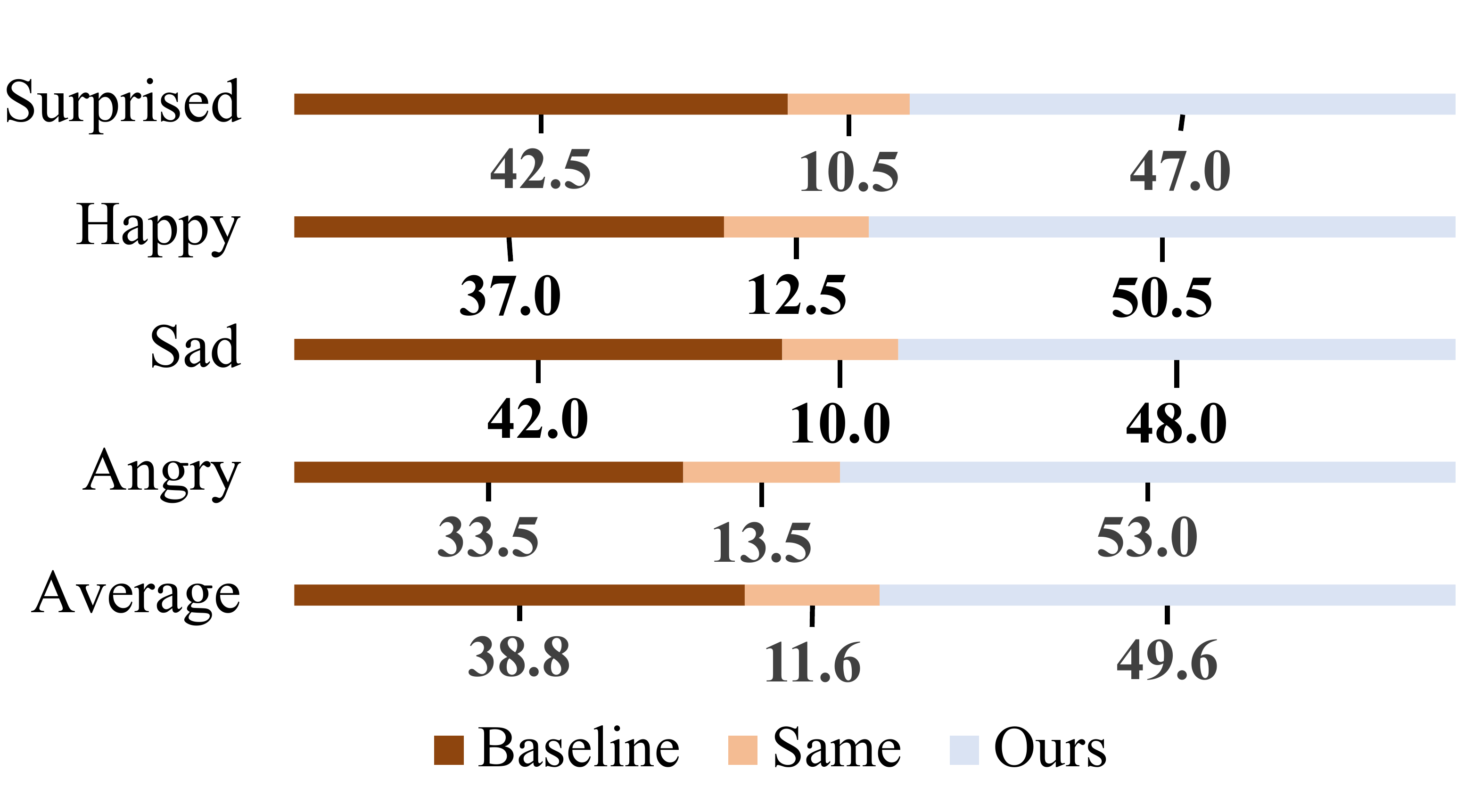}
  \caption{Preference test for emotion expressiveness. Both TTS models are trained on an imbalanced ESD dataset.}
  \label{fig:preference}

\end{figure}

The results are shown in Fig.~\ref{fig:preference}. 
As we can see, our model outperforms the baseline for all emotions and subjects are barely confused, which suggests that (1) our model's emotional expression is more clear than the baseline's; 
(2) despite sharing the same architecture with the original Reference Encoder in the baseline, the proposed Emotion Encoder is able to extract more effective and more robust emotional representations from an imbalanced dataset.  

\begin{figure}[t]
  \centering
  \includegraphics[width=0.65\linewidth]{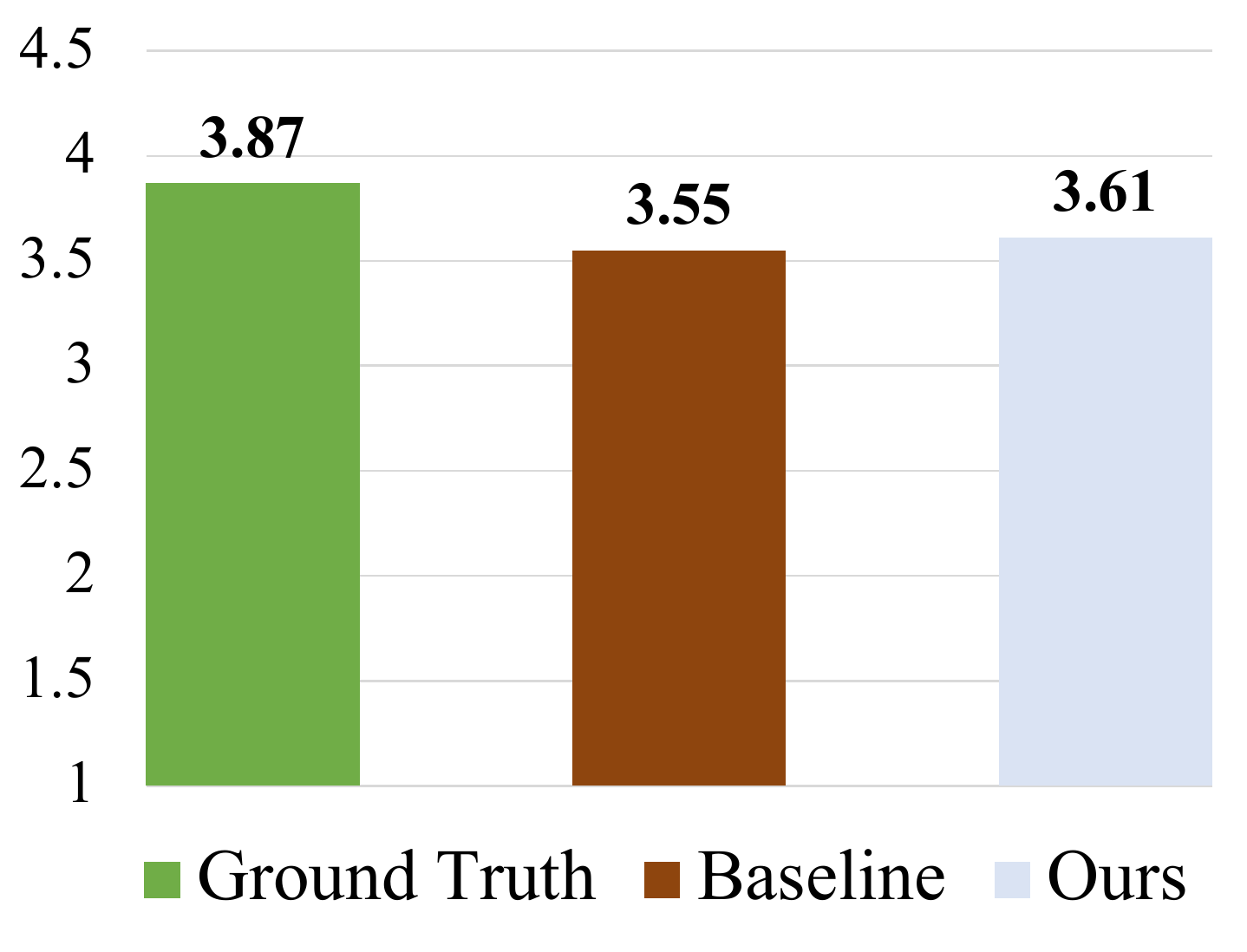}
  \caption{Naturalness MOS Scores.}
  \label{fig:mos}

\end{figure}

\subsubsection{Naturalness Test}

We further evaluate the naturalness of synthesized speech samples. Subjective measurement Mean Opinion Score (MOS) tests are conducted for this evaluation. During the test, subjects score the speech's naturalness on a scale of 1 (Bad) to 5 (Excellent).

The results are shown in Fig.~\ref{fig:mos}. 
As we can see, both models are close to the results of the ground-truth speech samples. 
Moreover, despite using the same architecture, our model slightly outperforms the baseline, 
indicating that our emotional representations might also bring benefits for the synthesis of natural speech.

\section{Conclusion}

In this paper, we propose an Emotion Extractor to extract speech emotional representations from imbalanced data.
The extracted representations are generalizable and effective. And we apply them to improve the performance of SER and Emotional TTS models.
The experimental results show that 
(1) for the SER task, our model surpasses the state-of-the-art baseline on three imbalanced datasets; 
(2) the learned representations benefit the TTS model and enable it to generate more expressive speech, even when the training dataset is imbalanced.

\bibliographystyle{IEEEtran}
\bibliography{mybib}

\end{document}